\documentclass[12pt]{article}
\usepackage{epsfig}

\def\beq{\begin{equation}}
\def\eeq#1{\label{#1}\end{equation}}
\def\eeqn{\end{equation}}

\def\beqa{\begin{eqnarray}}
\def\eeqa#1{\label{#1}\end{eqnarray}}
\def\eeqan{\end{eqnarray}}

\let\bar=\overbar

\def\Dslash{\not{\hbox{\kern-4pt $D$}}}
\def\dslash{\not{\hbox{\kern-2pt $\del$}}}

\def\msb{{\bar{\ssstyle M \kern -1pt S}}}

\def\Title#1{\begin{center} {\Large {\bf #1} } \end{center}}

\begin{document}

\Title{Measurement of Neutral Current Coherent $\pi^0$ Production In The NOvA Near Detector
}

\bigskip\bigskip


\begin{raggedright}  

{\it Hongyue Duyang\index{Duyang, H.} \\
Department of Physics and Astronomy\\
University of South Carolina\\
Columbia, South Carolina, United States\\for the NOvA Collaboration}
\bigskip\bigskip
\end{raggedright}

\section{Introduction}

Neutrinos can coherently interact with the target nucleus via neutral current (NC) exchange and produce an outgoing $\pi^0$
which involves a very small momentum transfer to the target nucleus and no exchange of quantum numbers. 
The characteristic of neutral current coherent $\pi^0$ final state is a single, forward-going $\pi^0$, with no other pions or nucleons or vertex activity. 

We are interested in the NC coherent $\pi^0$ for two reasons.
First, coherent $\pi^0$ is an important contribution to the background of the long-baseline $\nu_e$ appearance oscillation measurement. 
In many neutrino detectors, the photons from $\pi^0$ decay are 
often difficult to separate from the shower induced by electrons in $\nu_e$-CC signal events.  
Measurement of coherent $\pi^0$ production provides a constraint on this $\pi^0$ background. 
Secondly, the coherent process has physics interest in its own right. It provides insight into the structure of the weak hadronic current, and a test of the Partially Conserved Axial Current (PCAC) hypothesis \cite{pcac}\cite{rs}\cite{bs}, which relates the coherent pion production cross section to pion-nucleus elastic scattering, and used in Rein-Sehgal model and many neutrino generators such as GENIE. 

\section{NOvA Near Detector and Neutrino Beam}

The NOvA experiment has two functionally identical detectors at the far site (Ash River, MN) and the near site (Fermilab) \cite{novadesign}.
The near detector (ND), designed to measure the neutrino flux previous to oscillation, also provides an excellent opportunity for measurement of neutrino interactions. 

The NOvA ND is a 290 ton tracking calorimeter formed by alternating vertical and horizontal planes constructed with polyvinyl chloride (PVC) cells filled with liquid scintillator.   
The target nuclei for neutrino interactions are dominantly carbon ($^{12}$C, 66.8\%) and hydrogen ($^1$H, 10.5\%) from the scintillator and chlorine ($^{35}$Cl, 16.4\%) from PVC cells, with small contribution from titanium ($^{48}$Ti), oxygen ($^{16}$O) and other nucleus.
Each plane is about $0.18$ radiation length, optimized for the measurement of EM showers, including the photon showers induced by the $\pi^0$ decay.

NOvA uses the neutrino beam generated by the Fermilab Main Injector, by colliding 120 GeV protons on a 1.2 m graphite target. 
The NOvA ND is 1 km from the neutrino source, 100 m underground, 14 mrad off from neutrino beam axis.
The neutrino flux seen in the NOvA ND is a narrow band beam peaked at 1.9 GeV, with 68\% of neutrinos between 1.1 and 2.8 GeV. 
Simulation shows the neutrino flux is dominantly $\nu_\mu$ (94\%), with a small contamination from $\nu_e$ and $\bar{\nu}_\mu$.

\section{Signal Selection and Background Constraint}
\begin{figure}
\begin{center}
\includegraphics[width=0.49\linewidth]{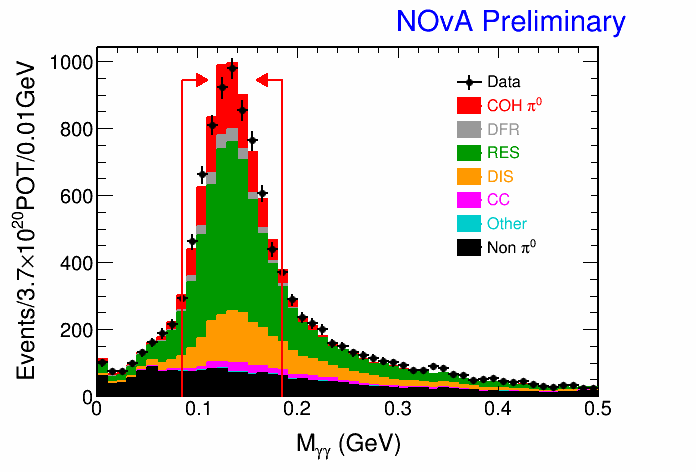}
\end{center}
\caption{Data and MC $\pi^0$ invariant mass distribution of the selected 2-prong NC $\pi^0$ sample. \label{mggnc}}
\end{figure}

In the NOvA ND, we select events with both photons from $\pi^0$ decay reconstructed as showers.
The photon showers are distinguished from background particles via log-likelihood functions based upon dE/dx information in both longitudinal and transverse direction of the showers. The invariant mass is calculated from the momenta of the reconstructed showers assuming both are photons: 
\begin{equation}
M_{\gamma\gamma} = \sqrt{2E_{\gamma1}E_{\gamma2}(1-cos\theta_{\gamma\gamma})}
\end{equation}
where $E_{\gamma1}$ and $E_{\gamma2}$ are the energy of the 2 photon showers and $\theta_{\gamma\gamma}$ is the opening angle. 
The invariant mass distribution shows good agreement between data and MC (Figure \ref{mggnc}) with the mass peak matching the known $\pi^0$ mass (134MeV).
Cuts are applied on the invariant mass distribution to choose only the peak region to reduce non-$\pi^0$ background. 
The coherent contribution to the NC $\pi^0$ sample is small compared to other interaction modes. 
The background dominantly comes from neutral current resonance (RES) and deep-inelastic scattering (DIS), with small contribution from diffractive (DFR) $\pi^0$ production and charged current interactions. 

\begin{figure}
\includegraphics[width=0.49\linewidth]{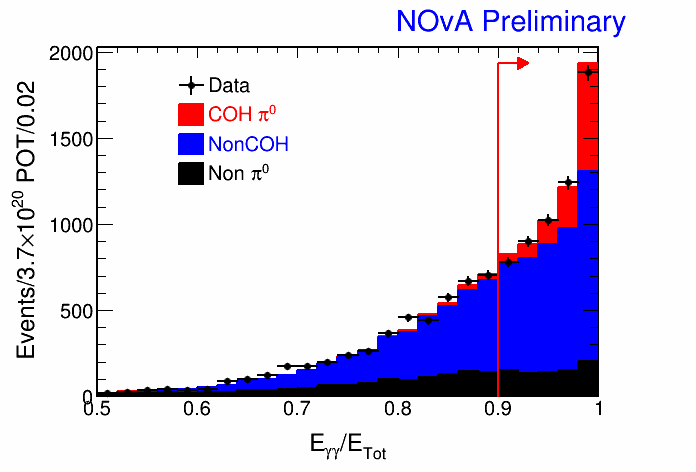}
\includegraphics[width=0.49\linewidth]{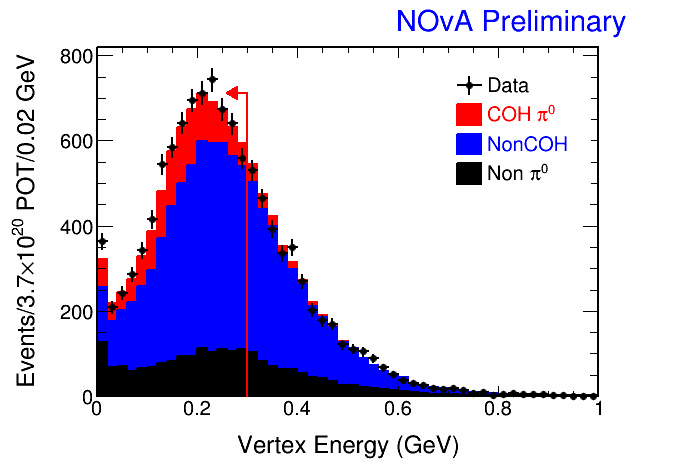}
\caption{$E_{\gamma\gamma}/E_{Tot}$ (top) and $E_{Vtx}$ (bottom) data/MC comparison. \label{caleratio_vtxe}}
\end{figure}

\begin{figure}
\begin{center}
\includegraphics[width=0.49\linewidth]{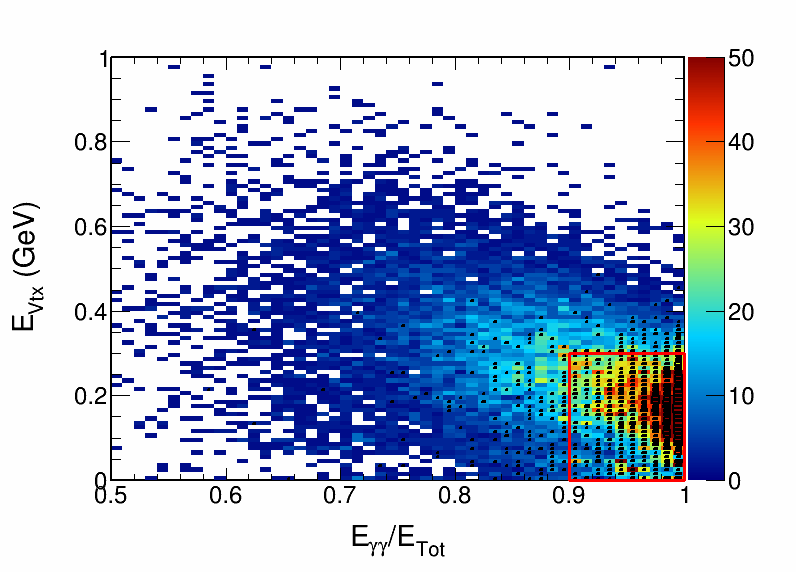}
\end{center}
\caption{$E_{\gamma\gamma}/E_{Tot}$ vs $E_{Vtx}$ in data (black box) and MC (color). The cut values are shown as the red box. The events inside the box is selected as the signal sample, and those outside the box is selected as the control sample. \label{cohsel2d}}
\end{figure}

To better control the background we further divide the NC $\pi^0$ sample into two independent sub-samples: a signal sample with most of the coherent signal for the cross-section measurement, and a control sample dominated by non-coherent $\pi^0$ for background constraining. 
Two variables are used for this purpose: the ratio of the shower energy to total event energy ($E_{\gamma\gamma}/E_{Tot}$), and the vertex energy ($E_{Vtx}$), defined as the energy on the first eight planes from the event vertex. 
Coherent interaction has one single $\pi^0$ in its final state with no other final state particles or additional vertex activity. 
The signal sample is therefore defined as events with most of their energy in the 2 photon-showers (large $E_{\gamma\gamma}/E_{Tot}$) and low vertex energy ($E_{vtx}$) to include most of the coherent signal and reduce background.
Rest of the events with extra energy other than photons or in the vertex region are defined as a control sample, dominated by non-coherent $\pi^0$s (RES and DIS).
The signal and control sample selection is illustrated in figure \ref{caleratio_vtxe} and \ref{cohsel2d} in $E_{\gamma\gamma}/E_{Tot}$ and $E_{Vtx}$ 1D and 2 space. 

\begin{figure}
\begin{center}
\includegraphics[width=0.49\linewidth]{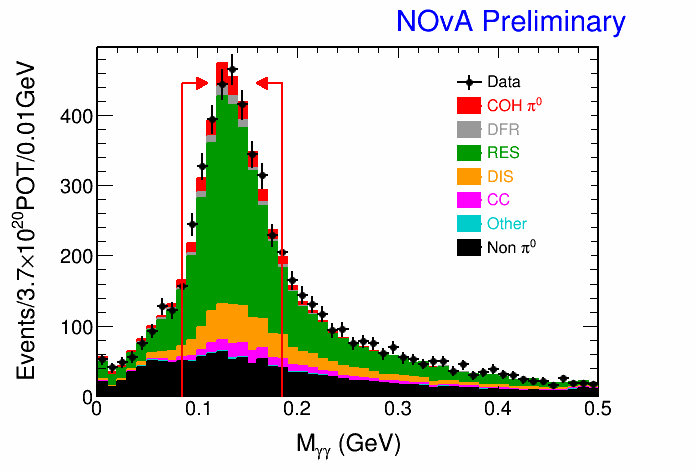} \\
\includegraphics[width=0.49\linewidth]{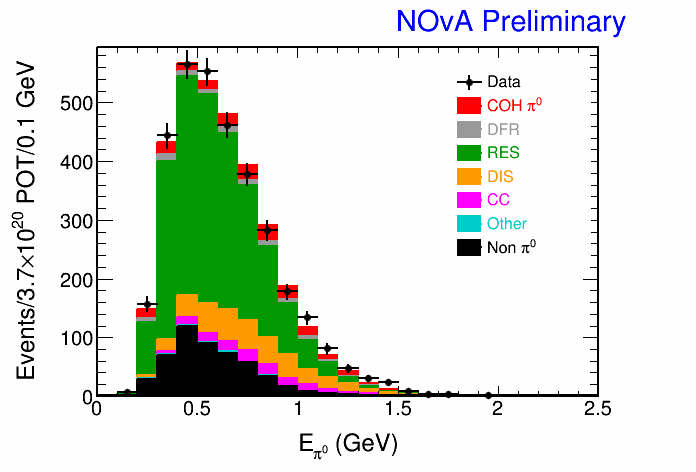}
\includegraphics[width=0.49\linewidth]{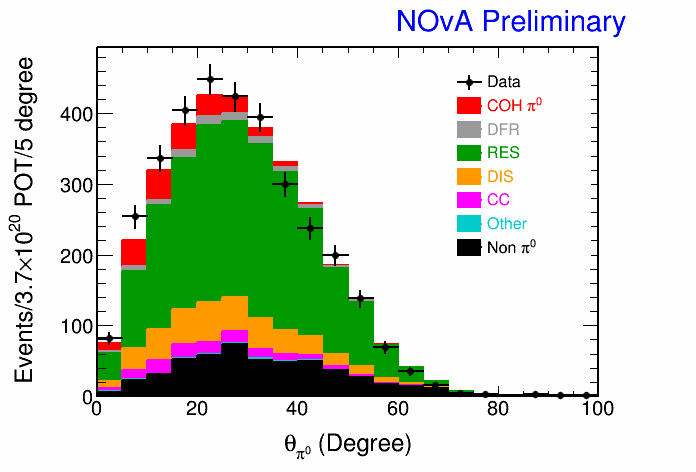}
\end{center}
\caption{$\pi^0$ energy and angle with respect to beam of the control sample events after the background fitting.
\label{egg_angle_ctr_norm}}
\end{figure}

\begin{figure}
\begin{center}
\includegraphics[width=0.49\linewidth]{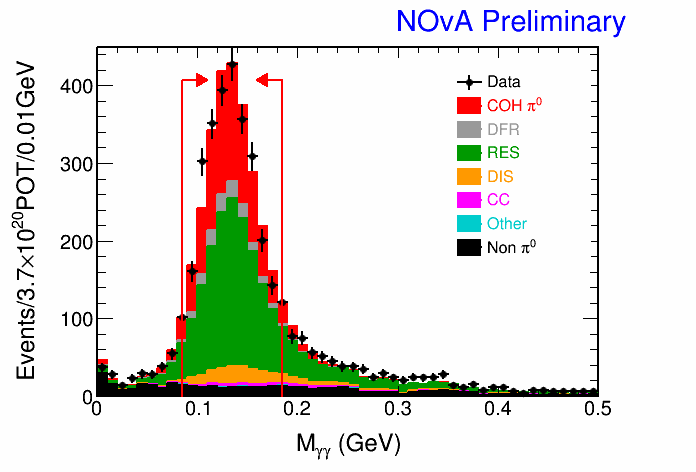} \\
\includegraphics[width=0.49\linewidth]{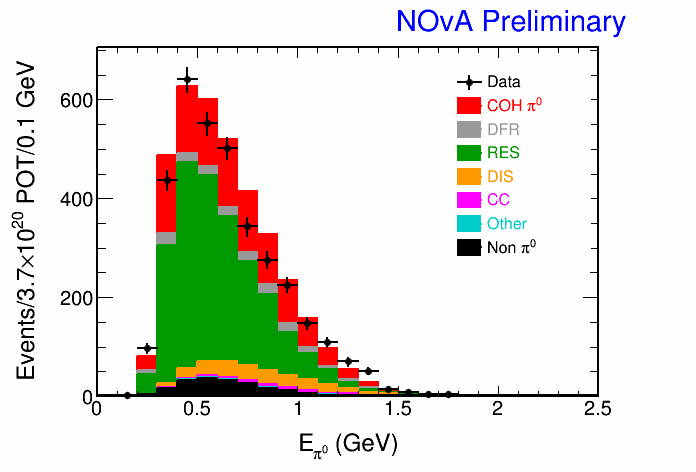}
\includegraphics[width=0.49\linewidth]{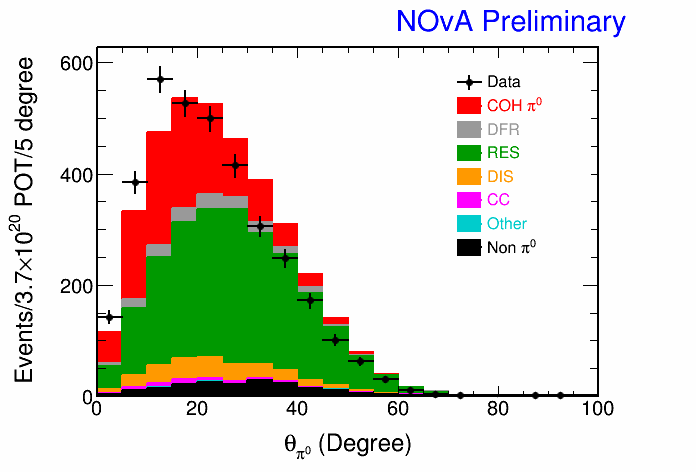}
\end{center}
\caption{$\pi^0$ energy and angle with respect to beam of the signal sample events after the background fitting.
\label{egg_angle_coh_norm}}
\end{figure}

The background is fit to control sample data by using RES and DIS as two templates.
Both the signal sample and control sample have non-coherent $\pi^0$ background dominated by NC RES and DIS.
In both samples, COH, RES and DIS show distinct distributions from each other in the $\pi^0$ energy and angle ($\cos\theta$) 2D space. 
Also the control sample RES/DIS has a very similar distribution to the signal sample RES/DIS.
The RES and DIS backgrounds in the signal sample are normalized according to the template fitting result.  
To further reduce the background, we define the coherent region in the 2D $\pi^0$ energy and $\cos\theta$ space.
The normalized background ($N_{Bkg, norm}$) in this coherent region is then subtracted from data ($N_{Data}$) to get the number of raw signal events ($N_{sig, raw}$). 
$987.4\pm67.3$(stat.) coherent signal events are observed. 
The invariant mass, energy and angle of the $\pi^0$s in control sample and signal sample after the fit is shown in figure \ref{egg_angle_ctr_norm} and \ref{egg_angle_coh_norm}.

\section{Systematic Uncertainty}
The systematic uncertainty for this analysis comes from calorimetric energy scale,  background modeling,  coherent signal modeling, 
detector response to photon showers,
detector simulation, particles entering the detector from the interaction in the rock surrounding the detector, and the simulation of neutrino flux.
Data-driven methods are used wherever possible to reduce the uncertainties. 
The calorimetric energy scale is constrained by the $\pi^0$ invariant mass distribution.
The background-related uncertainty is constrained by control sample data through the template fit method.  
We vary the background modeling parameters within $\pm\sigma$ according to GENIE and repeat the template fit. 
The uncertainty from each parameter is defined as the maximum deviation from the nominal value. 
Coherent modeling also introduces uncertainty via the efficiency correction. 
This effect is evaluated by varying the modeling parameters in the RS model: axial mass ($M_{A}$, $\pm50\%$) and nucleus radius ($R_{0}$, $\pm20\%$). 
To check the simulation of detector's response to photon showers, we identify the bremsstrahlung showers induced by rock muons and remove the muons to create a single photon control sample from data and MC \cite{cosmicmr}. The sample is subject to the same selection cuts as the $\pi^0$ photons and the uncertainty is evaluated as the difference between data and MC in selection efficiency.
Lastly, the neutrino flux uncertainty is constrained by external hadron production data. 
The systematic sources and uncertainties are summarized in table \ref{uncertainty}.
The total systematic uncertainty is determined to be 16.6\%.

\begin{table}
\begin{center}
\caption{List of systematic and statistical uncertainties. \label{uncertainty}}
\begin{tabular}{||c|c||}
\hline 
Source & $\delta (\%)$ \\ 
\hline 
Calorimetric Energy Scale & 3.4\\ 
Background Modeling & 10.0\\ 
Control Sample Selection & 2.9\\ 
Coherent Modeling & 3.7\\
Photon Shower Respond & 1.1\\ 
Rock Event & 2.4\\
Detector Simulation & 2.0\\ 
Flux  & 9.4\\ 
\hline 
Total Systematics& 15.3\\ 
\hline 
Signal Sample Statistics& 5.3\\ 
Control Sample Statistics& 4.1\\ 
\hline 
Total Uncertainty& 16.7\\
\hline 
\end{tabular} 
\end{center}
\end{table}

\section{Cross-Section Result}
The cross section is calculated using equation:
\begin{equation}
\sigma = \frac{N_{Data, selected} - N_{Bkg, norm}}{\epsilon\times N_{Target} \times \phi}
\end{equation}
where $N_{Data, selected} $ and $N_{Bkg, norm}$ are the number of data and normalized background MC in the selected coherent region of the signal sample,
 $\epsilon$ is the efficiency of coherent signal selection calculated from MC, 
 $N_{Target}$ is the number of target nucleus in the fiducial volume,
 and $\phi$ is the neutrino flux. 
 
\begin{figure}
\begin{center}
\includegraphics[width=0.6\linewidth]{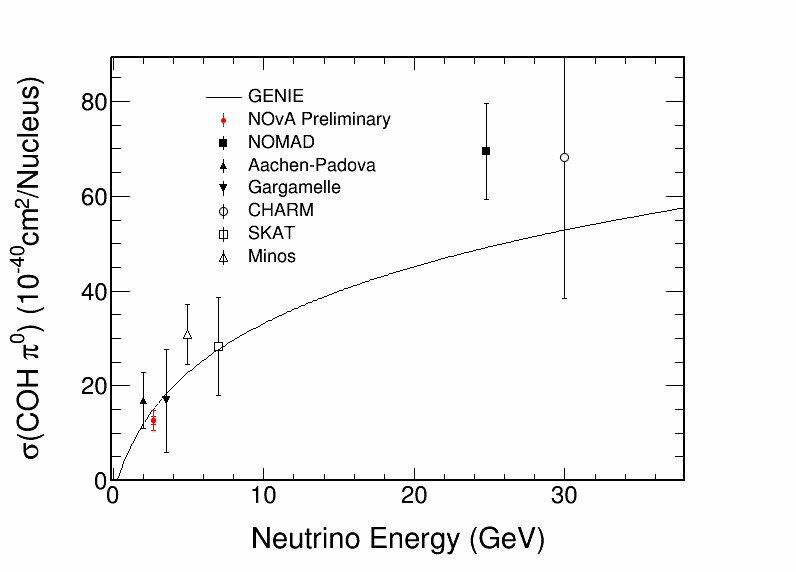}
\caption{Cross section of the NOvA coherent $\pi^0$ measurement comparing to world data (Aachen-Padova\cite{AP}, Gargamelle\cite{Gargamelle}, CHARM\cite{charm}, SKAT\cite{SKAT}, 15' BC \cite{BC}, NOMAD\cite{nomad}, MiniBooNE\cite{miniboone}, SciBooNE\cite{sciboone}, and Minos\cite{minos}) and RS model prediction from GENIE. Both statistical uncertainty, and statistical plus systematic uncertainty are shown as error bars. All results are scaled to carbon target by a factor of $(A/12)^{2/3}$ following Berger-Seghel model \cite{bs}, where A is the average atomic number of the certain experiment, to compare with each other and with the GENIE spline.
\label{xsec}}
\end{center}
\end{figure}

The measured cross section is $\sigma = 14.0\pm0.9(stat.)\pm2.1(syst.) \times 10^{-40}cm^2/nucleus$ at the average neutrino energy of 2.56 GeV. 
The effective atomic number $A = 13.8$ is calculated as the average atomic number of each element in the detector fiducial volume weighted by its contribution in total number of nucleus.
Figure \ref{xsec} shows the cross-section measurement of NOvA together with world measurements and GENIE prediction. 

\section{Summary}
To summarize, we have conducted a measurement of neutrino-induced coherent $\pi^0$ production using high statistics NOvA data. A data-driven method is developed to constrain the non-coherent background. 
The total uncertainty is 16.7\% including systematic and statistical uncertainties. 
This is one of the most precise measurement of coherent $\pi^0$ production in the world.\\
\bigskip

\section*{Acknowledgments}
NOvA is supported by the US Department of Energy; the US National Science Foundation; the
Department of Science and Technology, India; the European Research Council; the MSMT CR,
Czech Republic; the RAS, RMES, and RFBR, Russia; CNPq and FAPEG, Brazil; and the State
and University of Minnesota. We are grateful for the contributions of the staffs of the University
of Minnesota module assembly facility and NOvA FD Laboratory, Argonne National Laboratory,
and Fermilab. Fermilab is operated by Fermi Research Alliance, LLC under Contract No. DeAC02-07CH11359
with the US DOE.

\def\Discussion{
\setlength{\parskip}{0.3cm}\setlength{\parindent}{0.0cm}
     \bigskip\bigskip      {\Large {\bf Discussion}} \bigskip}
\def\speaker#1{{\bf #1:}\ }
\def\endDiscussion{}

 
\end{document}